\documentclass{article}
\usepackage{graphicx} 
\usepackage{natbib}
\usepackage{tikz}
\usetikzlibrary{arrows.meta}
\tikzset{%
	>={Latex[width=2mm,length=2mm]},
	base/.style = {rectangle, rounded corners, draw=black,
		minimum width=2cm, minimum height=1cm,
		text centered, font=\footnotesize, align=left},
	activityStarts/.style = {base, fill=blue!5},
	startstop/.style = {base, fill=blue!1},
	activityRuns/.style = {base, fill=red!5},
	squarednode/.style={base, fill=red!13},
	process/.style = {base, fill=orange!1}
}
\usepackage{xcolor}
\usepackage{ragged2e}
\usepackage{blindtext}
\usepackage{pdfpages}
\usepackage{lmodern}
\usepackage{enumitem}
\usepackage{amsmath}
\usepackage{amssymb}
\usepackage{amsthm}
\usepackage{adjustbox}
\usepackage{blindtext}
\usepackage{import}

\usepackage{titlesec} 
\usepackage{amsmath}
\usepackage{lmodern}

\usepackage{titlesec}
\usepackage{eurosym}
\usepackage{geometry}
\usepackage{stackengine}
\usepackage{listings}
\usepackage{float}
\usepackage{array}
\usepackage{graphicx}
\usepackage{stmaryrd}
\usepackage{framed}
\usepackage{tcolorbox}
\usepackage{adjustbox}
\usepackage{placeins}
\usepackage[font=small,labelfont=bf,justification=centering]{caption}
\usepackage{multirow}
\usepackage{epstopdf}
\usepackage{subcaption}

\usepackage{pgf,tikz,pgfplots}
\pgfplotsset{compat=1.15}
\usepackage{mathrsfs}
\usetikzlibrary{arrows}
\usepackage{chronosys}
\usepackage{hyperref}
\usepackage[nameinlink]{cleveref}
\usepackage{pdfpages}
\usepackage{setspace}
\usepackage{hyphenat}
\hypersetup{
	colorlinks,
	linkcolor={red!50!black},
	citecolor={blue!50!black},
	urlcolor={blue!50!black},
}

\usepackage{siunitx}
\renewrobustcmd{\bfseries}{\fontseries{b}\selectfont}
\renewrobustcmd{\boldmath}{}
\newrobustcmd{\B}{\bfseries}



\usepackage{footnotebackref} 
\usepackage[hang,bottom,]{footmisc}
\usetikzlibrary{positioning,shadings}
\usetikzlibrary{arrows}

\usepackage{dcolumn}
\newcolumntype{d}[1]{D..{#1}} 

\usepackage{tabularx}
\newcolumntype{L}[1]{>{\raggedright\arraybackslash}p{#1}} 
\newcolumntype{C}[1]{>{\centering\arraybackslash}p{#1}} 
\newcolumntype{R}[1]{>{\raggedleft\arraybackslash}p{#1}} 

\usepackage{booktabs}
\usepackage{multirow}
\usepackage{pdflscape}
\usepackage{fancyhdr}
\usepackage{afterpage}
\usepackage{longtable}
\usepackage{afterpage}
\usepackage{pifont}
\usepackage{lipsum}

\makeatletter
\def\ifGm@preamble#1{\@firstofone}
\appto\restoregeometry{%
  \pdfpagewidth=\paperwidth
  \pdfpageheight=\paperheight}
\apptocmd\newgeometry{%
  \pdfpagewidth=\paperwidth
  \pdfpageheight=\paperheight}{}{}
\makeatother

\makeatletter
\def\chron@selectmonth#1{\ifcase#1\or January\or February\or March\or April\or%
  May\or June\or July\or August\or September\or October\or November\or December\fi}
\makeatother

\usepackage{etoolbox}

\usepackage[toc,page,header]{appendix} 
\usepackage{minitoc} 
\mtcsetfont{parttoc}{chapter}{\normalsize\bfseries\sffamily}           
\mtcsetfont{parttoc}{section}{\normalsize\sffamily}            
\mtcsetfont{parttoc}{subsection}{\normalsize\sffamily}         
\mtcsettitlefont{parttoc}{\LARGE\bfseries\sffamily}            

\title{Bibliometrics}
\author{ }
\date{\today}

\begin{document}

 \date{{\footnotesize This version: \today.}}


\title{\vspace{-2cm}
The Rise of Health Economics:\\Transforming the Landscape of Economic Research}

\author{Lorenz Gschwent \;\; Björn Hammarfelt  \;\;   \\  
	Martin Karlsson \;\; Mathias Kifmann\thanks{{\footnotesize Gschwent: University of Duisburg-Essen, lorenz.gschwent@uni-due.de;  Hammarfelt: University of Borås,  bjorn.hammarfelt@hb.se; Karlsson: CINCH, University of Duisburg-Essen, martin.karlsson@uni-due.de; Kifmann: University of Hamburg, mathias.kifmann@uni-hamburg.de.}}}

\maketitle

\thispagestyle{empty}

\vspace{-0.5cm}

\begin{abstract}

\doublespacing

\noindent 
This paper explores the evolving role of health economics within economic research and publishing over the past 30 years. Historically largely a niche field, health economics has become increasingly prominent, with the share of health economics papers in top journals growing significantly. We aim to identify the factors behind this rise, examining how health economics contributes to the broader economic knowledge base and the roles distinct subfields play. Using a combination of bibliometric methods and natural language processing, we classify abstracts to define health economics. Our findings suggest that the mainstreaming of health economics is driven by innovative, high-quality research, with notable cyclicality in quality ratings that highlights the emergence and impact of distinct subfields within the discipline.

\end{abstract}

\vspace{0.5cm}



\newpage
\onehalfspacing
\setcounter{page}{1} 

\section{Introduction}

The purpose of this article is to explore the evolving role of health economics within general economic research and publishing over the past 30 years. Historically, despite seminal contributions from scholars like \citet{arrow1963uncertainty} on medical care economics, \citet{newhouse1970toward} on hospitals, and \citet{grossman1972concept} on health production and the demand for health, health economics remained largely a niche field. It rarely featured in the most prestigious general-interest economics journals.

\begin{figure}[H]
\centering
      \includegraphics[width=0.6\linewidth]{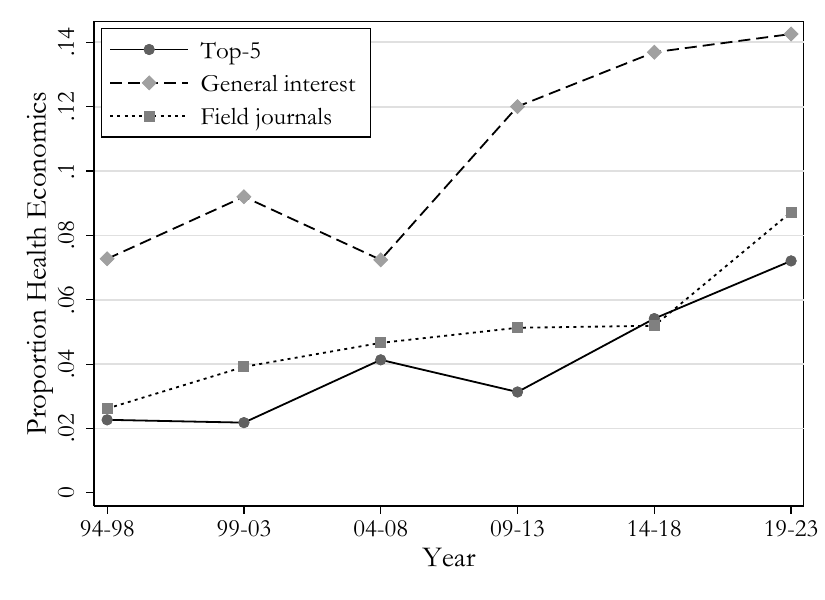}
      \caption{Proportion of Health Economics Papers by Type of Outlet}    \label{fig:HEPapers}
      	\scriptsize{}\textsc{Note.---} `Top-5' includes AER, JPE, QJE, REStud, and Econometrica; `General interest' includes REStat, AEJ:Applied, AEJ:Economic Policy, EJ, JEEA, RAND Journal, JHR; `Field journal' includes J Dev Econ, J Econ Growth, J Econometrics, J Int Econ, JOLE, JPubE. Papers are labeled as ``health economics'' based on the classification algorithm presented in Section \ref{papclass}.
\end{figure}

Over recent decades, the situation has changed dramatically. Health economists have increasingly entered the mainstream, and leading economists have increasingly turned their attention to health-related topics. Health economics is one of the fastest growing fields within economics \citep{bornmann2024recent}. This trend is illustrated in Figure \ref{fig:HEPapers}, showing the rising proportion of health economics papers in various types of journals. Between the mid-1990s and 2020, the share of health economics papers in ``top-5'' journals grew from 2\% to 6\%, while their presence in other general interest journals doubled from 7\% to 14\%. Even journals focused on distinct though related fields (labor, development, public economics) saw a quadrupling of health economics papers during this period \citep[cf.][]{mitra2020development}.

This paper aims to identify the factors behind the rising prominence of health economics in academic publishing. Specifically, we address how health economics contributes to the broader economic knowledge base and the roles that distinct subfields within health economics play in this process.

Our methodology involves multiple approaches. Firstly, we define what constitutes health economics, a complex task in bibliometrics. We examine the traditional method of classification based on JEL codes, discussing its pros and cons, and then develop an alternative approach leveraging advances in natural language processing (NLP). Using \texttt{RoBERTa} \cite{liu2019roberta}, a transformer-based large language model, we classify abstracts to determine if they belong to health economics. Our final classification combines two independent approaches, one representing the idea that health economics is what the dedicated field journals publish (as opposed to journals from other fields), and one representing the idea that health economics is research conducted by health economists. We show that this combined approach outperforms either method alone. In contrast to a classification based on JEL codes, our approach delivers a unique, exhaustive, and mutually exclusive classification of papers.

Second, we seek to understand whether the rise of health economics research is attributable mainly to disruptive, innovative research that represents a break from the past and leaves a lasting impact -- or rather to conventional research which excels in the current state of the art. In this part we rely on a method proposed by \cite{kelly2021measuring} to measure the \textit{novelty}, \textit{impact}, and \textit{quality} of patents, based on their similarity with earlier and later patents. `Novelty' captures low backward similarity, `impact' captures high forward similarity, and `quality' is a combination of novelty and impact. Adapting their approach to the specifics of academic publishing, we find that health economic papers consistently rate higher on `quality' than papers from other fields, suggesting that their rise is driven by innovation rather than conformity. Additionally, we observe cyclicality in the `quality' ratings of health economics research, distinct from other economic fields. We are able to show that distinct subfields contribute to these ``quality'' booms -- and leave a lasting impact on economic research.


The paper is organised as follows. The next section gives an overview of our analysis sample and presents the methods we use for paper classification and for the rating of novelty and impact of papers. Section \ref{sec:Results} presents the main empirical results and section \ref{sec:Conclusion} concludes.

\section{Data and Methods}

\subsection{Analysis Sample}

The analysis sample, and the data sample that provides the training data for classifications, consists of the universe of articles (excluding non-research contributions) published in 25 prominent economics journals over the 1994-2023 period. The sample of journals was selected according to the following principles:\begin{enumerate}
    \item Five journals identified as the most prominent field journals of health economics according to \citet{hammarfelt2023nordic}: JHE, HE, AJHE, EJHE, IJHEM.
    \item Prestigious journals with a remit that includes health: AER, JPE, QJE, ECMA, ReStud, AEJ: Applied, AEJ: Ec Policy, RESTAT, JEEA, EJ, JHR, JPubE, RAND, J Dev Econ, JoLE.
    \item Field journals for non-health fields: AEJ: Macro, JoLE, JPubE, J Econ Growth, J Dev Econ, Theoretical Economics, Journal of International Economics, Journal of Econometrics.
\end{enumerate}

The idea underlying this selection is that the analysis sample should include all potential outlets for health economic research that are at least as prestigious within the profession as the top field journal JHE.\footnote{The selection of journals in this part was based on an international survey of health economists, who provided answers with a great degree of agreement.} This is the motivation for including category (2). In addition, a journal-based classification of abstracts requires including journals from other fields, in particular fields that are related to health economics (so as to avoid false positives). This is the motivation behind category (3).

\subsection{Paper Classification}\label{papclass}

Operationalizing fields and specialties in bibliometrics is a challenging task. The most common approach is to base it on specific journals due to its straightforwardness \citep[cf.][]{mitra2020development,hammarfelt2023nordic}. However, this approach is not viable in our case since we aim to study the impact of health economics on general academic publishing in economics. Instead, bibliometric studies of economics often rely on the JEL classification of papers. One challenge is that the JEL classification aims to categorize both economists and their output for a wide range of stakeholders, including researchers, publishers, recruiters, and various external entities. The codes have resulted from numerous influences, external demands, and differing visions of the discipline \citep{cherrier2017classifying}. Therefore, the classification may deviate from what is optimal from a bibliometric point of view. For example, the JEL codes for a single paper may span multiple fields. Besides, the ordering of JEL codes is inconsistent over time \citep[cf.][]{angrist2020inside}, and JEL codes for a certain field may be added despite having negligible content related to that field \citep{wagstaff2012four}. \citet{kosnik2018survey} reports that even though the various JEL codes assigned to papers broadly reflect their contents, there is a striking disagreement between editors and authors regarding the appropriate JEL classification of papers.

Despite the known issues with JEL codes, researchers rarely explore alternative approaches. Some alternatives include using keywords to define a field \citep{geiger2017rise} or identifying the corpus of significant papers based on review articles \citep{braesemann2019behavioural}. A recent study in economic history incorporated information from abstracts and main texts in addition to JEL codes, highlighting the limited reliability of the JEL classification as a key motivation \citep{cioni2023economic}.

In this paper, we utilize recent advances in natural language processing (NLP) to classify economics publications into health economics or other sub-fields based on their titles and abstracts. Specifically, we employ RoBERTa, a large language model (LLM) based on the transformer architecture, known for its high performance in text classification tasks \citep{liu2019roberta, vaswani2023attention}. While the application of LLMs in economics has been limited due to concerns about performance on longer texts and interpretability \citep{ash2023text}, recent studies have successfully used LLMs for classification tasks in economics, such as job postings \citep{hansen2023remotework} and social media posts \citep{gehring2023analyzing}.

To classify health economics papers, we created two labeled datasets. Each dataset allows us to approach the classification of health economics publications as a binary classification problem with the label "1" representing health economics and the label "0" representing other economics publications. The first dataset, termed \textit{journal-based classification}, labels papers published in health economics journals as health economics and those in other fields (category 3 above) as non-health economics. This dataset includes 6,339 health economics papers out of a total of 19,434 papers.

Defining a field based on field journals is common in bibliometric research \citep[cf.][]{hammarfelt2023nordic}. However, one potential concern is that titles and abstracts in general-interest journals may be different in style to titles and abstracts in field journals, leading to poor external validity. To address this, we consider a second approach, termed \textit{author-based classification}. This approach entails two steps: first we use the field journals (categories 1 and 3 above) to classify authors as either health economists or non-health economists, depending on whether more than 50\% of their publications are in health economics journals. We then turn to the general interest journals (category 2, excluding the field journals listed in category 3) and label papers authored by health economists as health economics. This dataset comprises 498 health economics papers out of a total of 12,464 papers.


On both datasets, we trained a classifier based on RoBERTa-large. The inputs for the classifiers are the concatenated titles and abstracts of the publications in our training samples. The first step of the transformer architecture consists of turning the raw text into a numerical representation, so-called text embeddings, via an encoder. These text embeddings can then be fed into the neural network structure of RoBERTa. On top of RoBERTa, we add a linear layer that reduces the dimensionality of the output from 1,024 to 2. This allows us to interpret the raw output as relative weights whether a paper is a health economics publication or belongs to a different field. Finally, we add a sigmoid layer, which applies a logistic function to the raw outputs, normalizing them between 0 and 1, thereby allowing us to interpret the final output of the classifier as probabilities. For training, we split our data into subsets with proportions of 0.7, 0.2, 0.1 for the training, validation, and testing data, respectively. We combined the two classification methods via ROC curves to enhance robustness and performance. The combined classifier was then used to predict health economics papers in a total sample of 36,314 papers.\footnote{This total includes the two training datasets and papers within category (2) that could not be labeled for training due to having either unlabeled authors or combinations of health and non-health authors.}



\subsection{Language Similarity Measures}

In order to gain a deeper understanding of how health economics papers differ from publications from other fields of economics, we adapt \cite{kelly2021measuring}'s methods for patents. We thus assess \textit{novelty}, \textit{impact}, and \textit{quality} of health economics publications. Instead of relying on novel word combinations and their relative occurrence (tf-idfs -- term frequency-inverse document frequency) like \cite{kelly2021measuring}, we use sentence embeddings from the transformer-based model, \texttt{sentence-t5-xl} \citep{ni2021sentencet5}. We compute vector representations for combined titles and abstracts of each paper, resulting in a 768-dimensional vector with each dimension capturing a different aspect of the meaning of the text. 

The general idea of the \cite{kelly2021measuring} approach is that high `novelty' is captured by low similarity to previous publications, while high `impact' is indicated by high similarity to future publications. `Quality' is simply the difference or ratio of past and future similarity. For our implementation, we define backward and forward similarity of a paper $i$ as follows:\begin{align}
BS_{i}&=\text{sim} \left(v_{i}, \Bar{u}_{i}\left(-5,-1\right) \right) \\
FS_{i}&=\text{sim} \left(v_{i}, \Bar{u}_{i}\left(1,5\right) \right),
\end{align}
\noindent where $v_i$ is the vector representation of paper $i$, and $\Bar{u}_{i}\left(a,b\right)$ is an average vector representation of all papers published $a$ to $b$ years after paper $i$, which, with some abuse of notation, is specified as:\begin{equation}
\Bar{u}_{i}\left(a,b\right)=\mathbb{E}\left[v_{j}\mid a\leq t_{j}-t_{i}\leq b\right],
\end{equation}
\noindent with the publication year of a paper $j$ denoted $t_{j}$. Finally, $\text{sim}(v,w)=\frac{vw}{\lVert v\rVert \lVert w\rVert}$ is the cosine similarity between $v$ and $w$.

Our indicator of \textit{quality} is defined as the difference between forward and backward similarity:\begin{equation}
Q_{i}=FS_{i}-BS_{i}.
\end{equation}

\noindent Hence, in this formulation, a paper of high quality brings innovation compared to the past but receives a following in later research. Like \cite{kelly2021measuring} we also decompose this `quality' measure into \textit{novelty} and \textit{impact} components. However, to avoid these components reflecting a paper's (dis-)similarity to a generic economics paper, we measure them relative to the paper's similarity to contemporaneous papers. Accordingly, we define \textit{novelty} as follows:\begin{equation}
N_{i}=PS_{i}-BS_{i},
\end{equation}

\noindent where $PS_{i}$ represents the similarity to papers published in the same year:\begin{equation}
PS_{i}=\text{sim} \left(v_{i}, \Bar{u}_{i}\left(0,0\right) \right).
\end{equation}

Finally, we define \textit{impact} analogously as $I_{i}=FS_{i}-PS_{i}$.

\section{Results}\label{sec:Results}

\subsection{Paper Classification}

This subsection provides an overview of the performance of our RoBERTa classifiers in classifying papers based on their content. Table \ref{tab:appperf} presents the statistics of how the classifiers performed in different training data samples, including journal-based and author-based samples, as well as in the combined dataset.

\begin{table}[H]
    \centering
    \caption{Combining Statistics -- Performance}\label{tab:appperf}
\scalebox{0.95}{    \begin{tabular}{lccccccccc}
    \toprule
  &\multicolumn{3}{c}{\textsc{Journal Sample} ($N=20,440$)}&\multicolumn{3}{c}{\textsc{Author Sample} ($N=7,053$)}&\multicolumn{3}{c}{\textsc{Combined} ($N=27,493$)}\\
\cmidrule(lr){2-4}\cmidrule(lr){5-7}\cmidrule(lr){8-10}
  Criterion& \textit{Sensitivity} & \textit{Specificity} & \textit{F1} & \textit{Sensitivity} & \textit{Specificity} & \textit{F1} & \textit{Sensitivity} & \textit{Specificity} & \textit{F1} \\
\midrule
\(p_J>0.500\)&       0.960&       0.969&       0.956&       0.481&       0.956&       0.441&       0.932&       0.964&       0.923\\
\(p_J>0.327\)&       0.961&       0.969&       0.956&       0.483&       0.956&       0.442&       0.933&       0.964&       0.923\\
\(p_A>0.500\)&       0.960&       0.953&       0.943&       0.708&       0.934&       0.513&       0.945&       0.946&       0.909\\
\(p_A>0.828\)&       0.943&       0.963&       0.941&       0.672&       0.948&       0.537&       0.926&       0.957&       0.912\\
\(p_C>0.498\)&       0.972&       0.967&       0.960&       0.581&       0.954&       0.501&       0.949&       0.962&       0.929\\

\bottomrule
    \end{tabular}}
\begin{minipage}{\textwidth} \vspace{0.1cm}   
\scriptsize{}\textsc{Note.}---The table reports the classification performance of the two RoBERTa classifiers. $p_J$ denotes the prediction probability of the \textit{journal-based} classifier, $p_A$ is the probability of the \textit{author-based} classifier, and $p_C$ represents the \textit{average} of the two. We report the standard test properties sensitivity and specificity along with the F1 score ($F1=\frac{2TP}{2TP+FP+FN}$), where $TP$ represents the true positive rate, $FP$ the false positive rate and $FN$ the false negative rate \citep{van1979information}.
\end{minipage}
\end{table}

The classifiers demonstrated excellent performance in the journal-based training data sample (left panel), with both sensitivity and specificity exceeding 0.95. However, sensitivity in the author-based sample (middle panel) was notably lower, reflecting the noisier nature of an author's identity as a signal for classification. Nevertheless, when the two samples are combined (right panel), the overall performance is very high in general. To ensure robust classification across all sub-samples, we use the global $F1$ score (rightmost column) as our main criterion. We find that averaging the test statistics and applying a cutoff of 0.498 yields the highest global F1 score. Thus, we implement this combined classifier in subsequent analyses.

\begin{figure}[H]
\centering
      \includegraphics[width=0.6\linewidth]{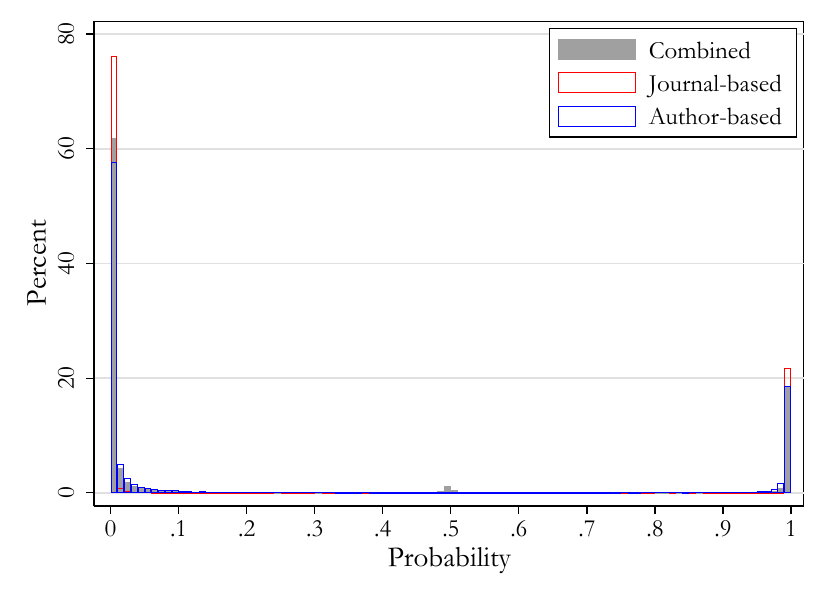}
      \caption{Distribution of Classification probabilities}    \label{fig:ClassHist}
      	\scriptsize{}\textsc{Note.---} Own calculations. Distribution of the RoBERTa classifiers' probabilities for a paper being a health economics publication in our main analysis sample ($N=38,116$).
\end{figure}

Figure \ref{fig:ClassHist} displays the distribution of classification probabilities within the entire analysis sample. It is evident that all three classifiers provide a clear classification in the vast majority of cases. As the histogram for our preferred combined classification shows, there are very few instances where the author-based and journal-based classifiers disagree: these are visible in the region around 0.5.

In order to give a flavour of how the classification works, Table \ref{tab:examples} provides examples of papers from six different groups, defined by how confidently they are classified as either health or non-health papers. Confidence is in this case operationalized as the the distance of the prediction probability from the cutoff. Additionally, we also visualize the transformer-based predictions via dimensionality reduction of the vector representations from sentence-t5-xl in Appendix \ref{app:mapping}.

\begin{table}[H]
    \centering
    \caption{Examples of Classifications by Degree of Confidence}\label{tab:examples}
      \renewcommand{\arraystretch}{1.5} 
   \scalebox{0.8}{ \begin{tabular}{p{8cm}cp{9.5cm}c}
    \toprule
    \textsc{Reference}&\textsc{Journal}&\textsc{Title}&\textsc{Score}\\
    \midrule
    \multicolumn{4}{l}{\textsc{I. Health Economics, High Confidence}}\\
    \midrule
    \citet{manning1996health} & JHE & Health insurance: The tradeoff between risk pooling and moral hazard  & 0.999 \\
    \citet{decker2005medicare} & JHR & Medicare and the health of women with breast cancer  & 0.999 \\
    \citet{coile2014recessions} & AEJ: EP & Recessions, Older Workers, and Longevity: How Long Are Recessions Good for Your Health?  & 0.999 \\
    \midrule
    \multicolumn{4}{l}{\textsc{II. Health Economics, Medium Confidence}}\\
    \midrule
   \citet{baines1996selection} & HE & Selection bias in GP fundholding & 0.998 \\
   \citet{rellstab2020kids} & JHE & The kids are alright -- labour market effects of unexpected parental hospitalisations in the Netherlands & 0.998 \\
   \citet{sabariego2011cost} & EJHE & Cost-effectiveness of cognitive-behavioral group therapy for dysfunctional fear of progression in cancer patients & 0.998 \\
   \midrule
    \multicolumn{4}{l}{\textsc{III. Health Economics, Low Confidence}}\\
    \midrule
   \citet{reisinger2019parallel} & JHE & Parallel imports, price controls, and innovation & 0.529 \\
   \citet{chari2017causal} & JDevEc & The causal effect of maternal age at marriage on child wellbeing: Evidence from India & 0.525 \\
   \citet{gundersen2008food} & JHR & Food stamps and food insecurity: what can be learned in the presence of nonclassical measurement error? & 0.519 \\
   \midrule
    \multicolumn{4}{l}{\textsc{IV. Not Health Economics, Low Confidence}}\\
    \midrule
   \citet{persico2021effects} & JHR & The effects of local industrial pollution on students and schools & 0.167 \\
      \citet{boomhower2019drilling} & AER & Drilling Like There's No Tomorrow: Bankruptcy, Insurance, and Environmental Risk & 0.152 \\
   \citet{knittel2004regulatory} & RESTAT & Regulatory restructuring and incumbent price dynamics: the case of US local telephone markets & 0.140 \\
   \midrule
    \multicolumn{4}{l}{\textsc{V. Not Health Economics, Medium Confidence}}\\
    \midrule
   \citet{gollin2016urbanization} & J Econ Growth & Urbanization with and without industrialization & 0.002 \\
   \citet{aradillas2022inference} & J Econometrics & Inference in ordered response games with complete information & 0.002 \\
   \citet{mintz2004income} & JPubE & Income shifting, investment, and tax competition: theory and evidence from provincial taxation in Canada & 0.002 \\
   \midrule
    \multicolumn{4}{l}{\textsc{VI. Not Health Economics, High Confidence}}\\
    \midrule
   \citet{alvarez2004habit} & J Econ Growth & Habit formation, catching up with the Joneses, and economic growth & 0.001 \\
   \citet{sampson2016dynamic} & QJE & Dynamic selection: an idea flows theory of entry, trade, and growth & 0.001 \\
   \citet{krusell2010temptation} & ECMA & Temptation and taxation & 0.001 \\

\bottomrule
    \end{tabular}}
      \renewcommand{\arraystretch}{1} 
  \begin{spacing}{1} 
    \scriptsize{}\textsc{Note.}---``High confidence'' corresponds to percentile 100 of confidence within each classification, ``Medium confidence'' corresponds to percentile 51 and ``Low confidence'' corresponds to percentile 5.
    \end{spacing}
\end{table}

\paragraph{Validation.} The results reported in Table \ref{tab:appperf} suggest that our classification has excellent internal validity. We also consider the agreement between our classification and other attempts at classifying economics papers. In this part, we draw upon the work of \citet{angrist2020inside}, who classified economics papers published in the 1970-2015 period into different fields based on their JEL codes. There are 23,906 papers in both datasets, of which 2,112 are classified as health economics according to our classifier.

\begin{figure}[H]
   \centering
   \begin{subfigure}{0.46\linewidth}
           \includegraphics[width=\linewidth]{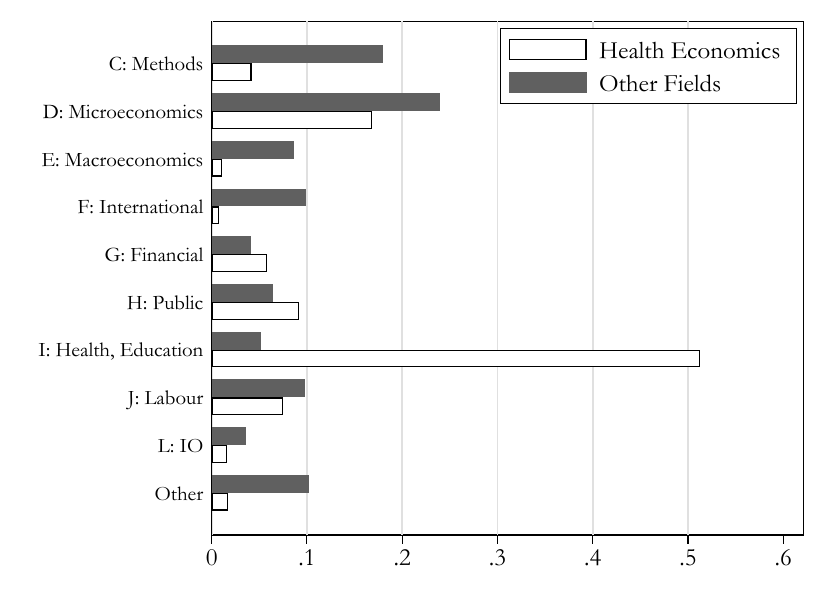}
   \caption{Primary JEL Code}\label{fig:JELs}\end{subfigure}
   \begin{subfigure}{0.46\linewidth}
      \includegraphics[width=\linewidth]{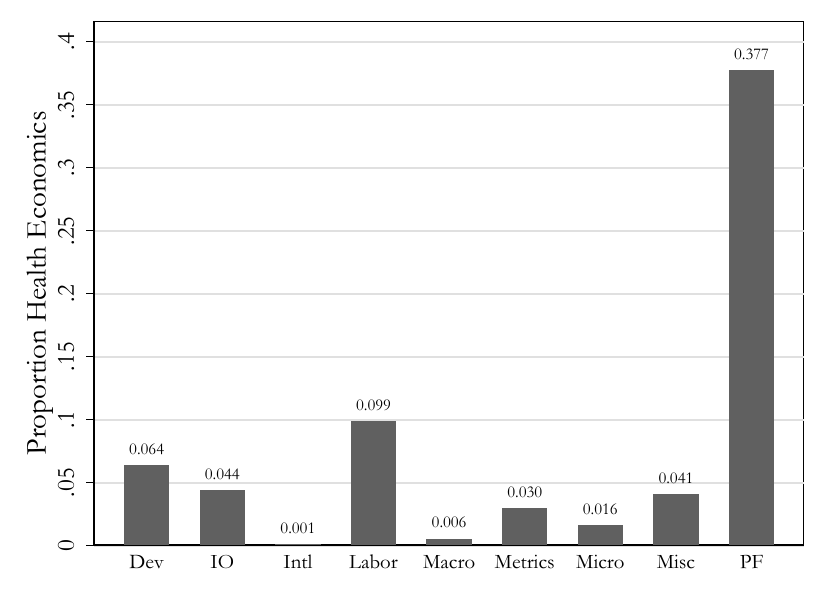}
   \caption{By field}\label{fieldsbar}\end{subfigure}
    \caption{Comparison of Classifications.} \label{fig:deepimpact}      	\scriptsize{}\textsc{Note.}---The left chart shows the distribution of papers over different JEL codes depending on their classification as health economics papers. The right chart shows the distribution of papers over different fields \citep[according to][]{angrist2020inside} depending on their classification as health economics papers.
\end{figure}

We report the distribution of papers across different fields, conditional on our classification, in Figure \ref{fig:deepimpact}. Figure \ref{fig:JELs} displays the distribution of health and non-health papers over primary JEL codes. As expected, category I, ``\textit{Health, Education, and Welfare},'' is the most prominent group among health economics papers. Interestingly, the categories G ``\textit{Financial Economics}'' and H ``\textit{Public Economics}'' also show an over-representation of health economics papers. The former is driven by category G22 ``\textit{Insurance, Insurance Companies, Actuarial Studies}'' (92 percent of papers in the group), and the latter is driven by the categories H51 ``\textit{Government Expenditures and Health}'' and H75 ``\textit{State and Local Government: Health, Education, Welfare, Public Pensions}'', which together account for 72 percent of papers in the group.

Figure \ref{fieldsbar} displays the proportion of health economics papers (according to our classification) within different fields of economics, as defined by \citet{angrist2020inside}. It is clear that the field ``public finance'' has by far the highest proportion; one-third of papers classified as public finance are health economics papers according to our classifier. In all other fields, the proportion of health economics papers is below 10 percent, with particularly low proportions in Macroeconomics and International Economics.

In Figure \ref{app:mapping} we provide an alternative type of validation; it is based on a reduction in dimensionality in the original text embeddings which makes it possible to visualise papers as co-ordinates in a two-dimensional space. We use a combined procedure of Principal Component Analysis (PCA) and t-distributed Stochastic Neighbour Embedding (t-SNE) to reduce the dimensionality of the sentence-t5-xl text embeddings. In Figure \ref{app:predclusters}, our predictions of Health Economics papers form a relatively coherent and homogeneous group within this mapping of economics papers. To provide some additional context, we report the result of hierarchical clustering \citep[cf.][]{campello2013hdbscan} of the dimensionality-reduced text embeddings. In the legend we highlight the five most common journals in each of these clusters. This further emphasises the relationship between health economics and labour, development, and public economics.

\begin{figure}[H]
    \caption{Locating Health Economics within Economics}\label{app:mapping}
    \centering
    \begin{subfigure}{.49\textwidth}
		\centering
        \includegraphics[width=\textwidth]{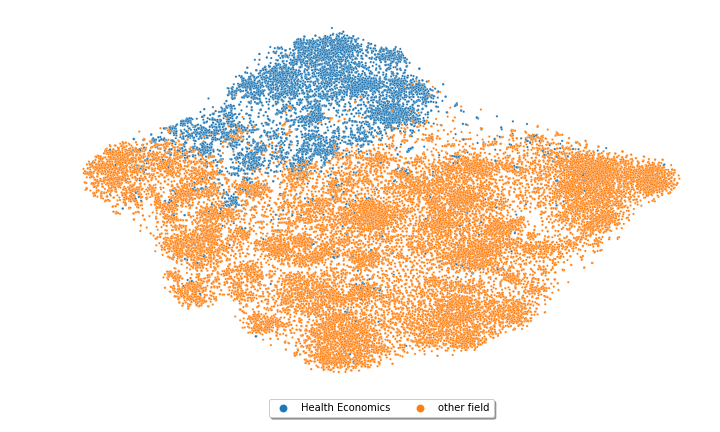}
		\caption{Health Economics predictions}\label{app:predclusters}
	\end{subfigure}
    \begin{subfigure}{.49\textwidth}
		\centering
        \includegraphics[width=\textwidth]{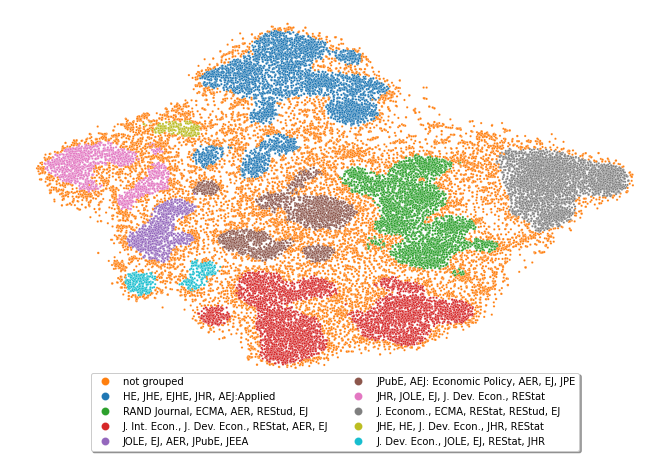}
		\caption{Clustering Economics publications}\label{app:fieldclusters}
	\end{subfigure}
\begin{minipage}{\textwidth}  
\scriptsize{}\textsc{Note.---} This Figure plots the dimensionality-reduced sentence-t5-xl text embeddings for every paper in our sample. To reduce the dimensions of the 768-dimensional sentence-t5-xl vectors, we employ a combination of PCA (reducing 768 dimensions to 50) and t-SNE (reducing 50 dimensions to 2). The nine distinct clusters in Panel (b) result from first clustering the two-dimensional text representations into 24 distinct clusters, which were then combined if they shared the most common journal.
\end{minipage}
\end{figure}

As a final validation exercise, we compare the outcome of our classification to the body of work studied by \citet{wagstaff2012four}. Their classification is based on a paper having a health JEL code in any position -- and they acknowledge that this is likely to generate false positives. We focus on the 300 papers over the 1969-2010 period that they identify as the most influential and study the classification of these papers in our data: the overlap is 73 papers. The mean prediction is 0.87 for our author-based classification and 0.89 for our journal-based classification; which suggests that the vast majority of the 73 papers would be classified as health economics by our classifier. Indeed, when we use our preferred combined classifier, all but 8 of the 73 papers get classified as health economics. The exceptions are fairly revealing and include \citet{acemoglu2001colonial} on the colonial origins of development; \citet{rodrik2004institutions} on the role of institutions, geography, and trade in determining income levels; and \citet{blau1999effect} on the effect of income on child development.

\subsection{Novelty -- Impact -- Quality}

\begin{figure}[H]
   \centering
   \begin{subfigure}{0.46\linewidth}
           \includegraphics[width=\linewidth]{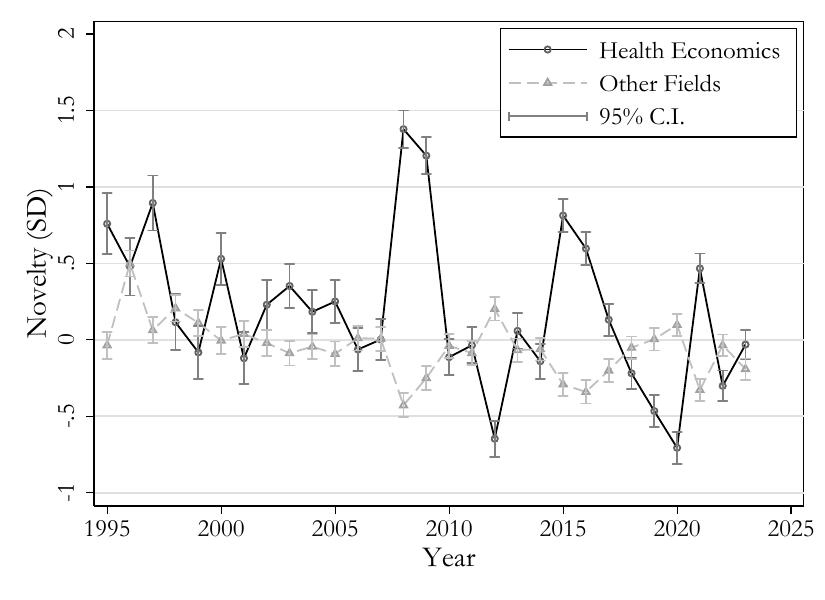}
   \caption{Novelty}\end{subfigure}
   \begin{subfigure}{0.46\linewidth}
      \includegraphics[width=\linewidth]{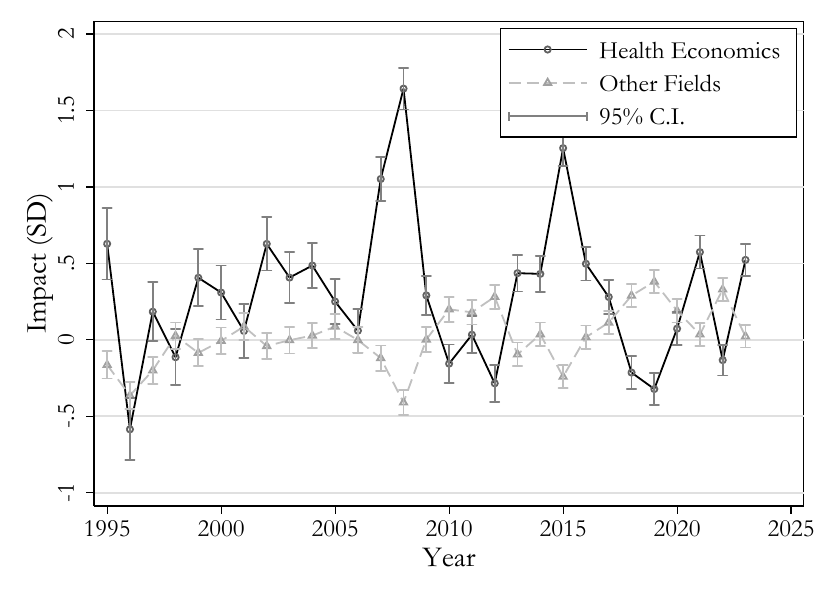}
   \caption{Impact}\end{subfigure}
    \caption{Novelty and Impact Ratings of Papers in Health Economics and Other Fields.} \label{fig:novimp}      	\scriptsize{}\textsc{Note.---} Estimates are based on the entire analysis sample ($N=35,208$) and include controls for journal fixed effects.
\end{figure}

In Figure \ref{fig:novimp} we compare how the `novelty' and `impact' indicators evolve over time for health and non-health publications. The figures suggests that health economics papers score higher on each indicator on average, but also that this advantage is concentrated in some peak years.

\begin{figure}[H]
   \centering
   \begin{subfigure}{0.46\linewidth}
           \includegraphics[width=\linewidth]{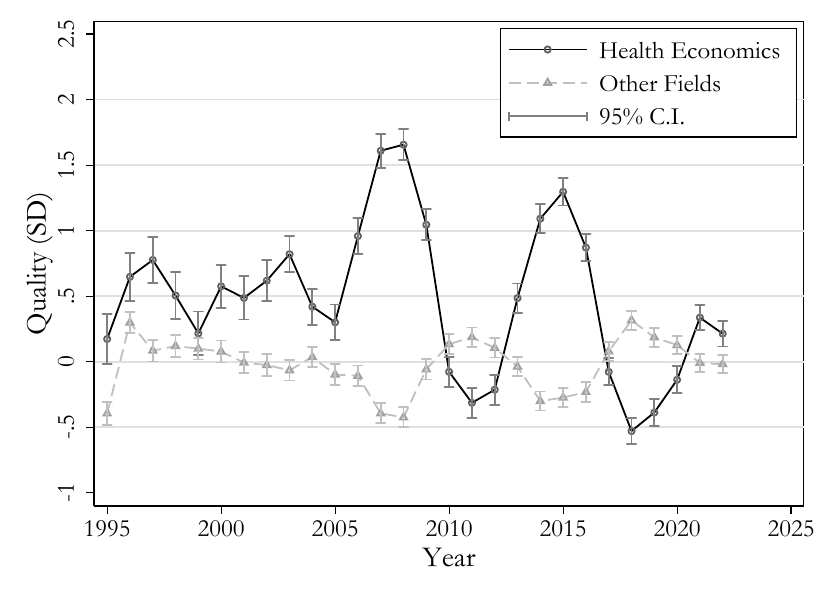}
   \caption{Health versus Non-Health}\end{subfigure}
   \begin{subfigure}{0.46\linewidth}
      \includegraphics[width=\linewidth]{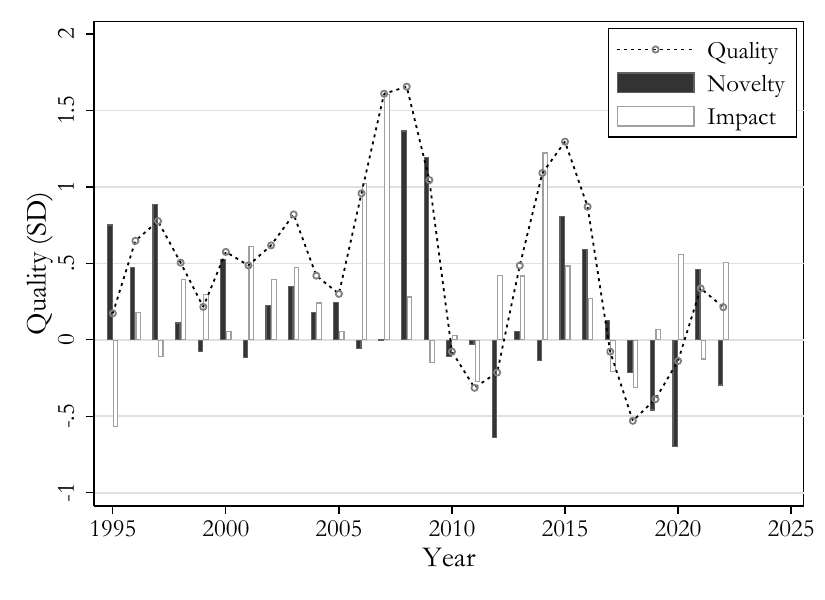}
   \caption{Novelty versus Impact}\label{novimp}\end{subfigure}
    \caption{Quality Ratings of Papers in Health Economics and Other Fields.} \label{fig:quality}      	\scriptsize{}\textsc{Note.---} In Figure \ref{novimp}, `novelty' and `impact' are measured in standard deviations of the `quality' score -- in order to ensure that the annual average of the `quality' score is correctly decomposed. Estimates are based on the entire analysis sample ($N=35,208$) and include controls for journal fixed effects.
\end{figure}

The corresponding graph for the combined `quality' measure is provided in Figure \ref{fig:quality}. Also in this case we observe a distinct advantage of health economics compared to other fields, particularly in certain years. To illustrate this, we provide the decomposition in Figure \ref{novimp}. Interestingly, both waves of increased `quality'---occurring in the 2006-2009 period and again in the 2014-16 period---are initially triggered by a surge in `impact', followed by increased `novelty'.

Examining each wave of increased `quality' scores as a whole, it proves challenging to discern clear patterns regarding the types of research influencing the improvement. However, a more detailed perspective, differentiating between the influence of `novelty' and `impact,' reveals a relatively coherent pattern. Notably, the substantial uptick in `quality' during 2007-08, primarily attributed to heightened `impact,' appears largely driven by emerging literature emphasizing the significance of the early life period for adult outcomes. The wave is thus to a great extent attributable to papers like \citet{bleakley2007disease,doyle2007child,ludwig2007does,birchenall2007escaping,black2007cradle} and \citet{chen2007long}. All of these papers score high on `impact', and all of them feature prominently in the review article on the ``fetal origins'' hypothesis published a few years later by \citet{almond2011killing}. The continuation of the wave in 2008-09 is instead mainly driven by a surge in `novelty'. This is to a great extent spurred by papers studying health care financing -- such as \citet{flores2008coping,grignon2008does,chi2008out,ariizumi2008effect} and \citet{wang2009impact}.

Regarding the 2014–16 wave, the segment characterized by high `impact' is notably propelled by research exploring the influence of financial incentives within health insurance schemes on the behaviors of both healthcare providers and patients. Noteworthy examples encompass \citet{clemens2014physicians,brown2014does,shigeoka2014effect} and \citet{chandra2014impact}. Conversely, the latter phase of this wave is driven by research scoring high on `novelty'. It includes a significant portion of papers focusing on empirical methods, such as \citet{jones2015healthcare,kreif2015evaluation,mccarthy2015putting} and \citet{armstrong2015asymptotically}. In summary, it appears that a focused investigation into the diverse fields contributing to fluctuations in measured `quality' could potentially reveal exactly what fields they are made up of.

\paragraph{Relationship to Citations.} A widely used measure of academic impact is the number of citations a paper generates. We assess the performance of our `novelty', `impact', and `quality' measures by examining how well they predict citations. First, we consider the general correlation patterns between these indicators and citations. Results from simple OLS regressions, where citations normalised by publication year are used as the dependent variable, are presented in Table \ref{tab:predcit}. In the left panel, we estimate the raw correlations between our three indicators and citations. A one-standard-deviation increase in novelty' is associated with a 3.8\% increase in citations, while a one-standard-deviation increase in `impact' is associated with a 4.3\% increase in citations. The difference between the coefficients is not statistically significant.

However, `novelty' and `impact' are negatively correlated, with a correlation coefficient of -0.49, which biases OLS estimates. Consequently, when we include both indicators in a single regression specification, the estimated coefficients almost double. In the fourth column, we estimate the coefficient of the combined `quality' indicator, which suggests that a one-standard-deviation increase in `quality' is associated with 8\% more citations.

\begin{table}[H]
    \centering
    \caption{Prediction of Citations}\label{tab:predcit}
    \begin{tabular}{lcccccccc}
    \toprule
  &\multicolumn{4}{c}{\textsc{Unconditional}}&\multicolumn{4}{c}{\textsc{Conditional on Journal}}\\
\cmidrule(lr){2-5}\cmidrule(lr){6-9}
& (1) & (2) & (3) & (4) & (5) & (6) & (7) & (8) \\
\midrule
Novelty                  &      0.0384***&               &      0.0777***&               &      0.0711***&               &      0.1352***&               \\
                         &     (0.012)   &               &     (0.013)   &               &     (0.011)   &               &     (0.013)   &               \\
Impact                   &               &      0.0431***&      0.0809***&               &               &      0.0551***&      0.1245***&               \\
                         &               &     (0.012)   &     (0.013)   &               &               &     (0.011)   &     (0.013)   &               \\
Quality                  &               &               &               &      0.0804***&               &               &               &      0.1318***\\
                         &               &               &               &     (0.012)   &               &               &               &     (0.012)   \\
\midrule Journal F.E.    &               &               &               &               &\(\checkmark\)   &\(\checkmark\)   &\(\checkmark\)   &\(\checkmark\)   \\
\midrule
N                        &      35,208   &      35,208   &      35,208   &      35,208   &      35,208   &      35,208   &      35,208   &      35,208   \\

\bottomrule
    \end{tabular}
\begin{minipage}{\textwidth}   
\scriptsize{}\textsc{Note.---}OLS regressions of year-normalised citation counts on the main analysis sample. In Figures \ref{app:binscatters} and \ref{app:binscatterswfe} in Appendix \ref{sec:app}, we report binned scatter plots for the univariate regressions in this Table.  Independent variables are measured in standard deviations. Standard errors in parentheses. * $p < 0.05$, ** $p < 0.01$, *** $p < 0.001$.
\end{minipage}
\end{table}

\noindent In columns (5) to (8), we estimate the correlation between our indicators and citations, conditional on the journal in which the papers were published. This changes the interpretation of the estimates, as the three indicators may affect the likelihood of being published in a particular journal. Indeed, we find that including journal fixed effects substantially magnifies the estimated coefficients, with factors ranging from 1.3 to 1.9.

We conclude that our three indicators carry meaningful information about the papers in the sense that they predict subsequent citations. We next turn to the question of whether health economics papers perform differently compared to non-health papers. This is of interest since there is a common conception that fields close to life sciences (such as health economics) tend to generate more citations than other fields. If this is the case, it would e.g. rationalize the trend we observe in Figure \ref{fig:HEPapers} to the extent that journal editors put increasing weight on citations.

Results are presented in Table \ref{tab:predcitH}, which has the same structure as Table \ref{tab:predcit} but includes interactions with our paper classifier. We can clearly refute the notion that health economics is treated more favorably than other fields. The intercept for health economics papers is lower than for other papers; however, this difference diminishes when we condition on the outlet. Additionally, we find that health economics papers are systematically less rewarded for ‘novelty,’ ‘impact,’ and ‘quality.’ Specifically, papers classified as health economics are associated with significantly lower coefficients for all three indicators. This finding remains remarkably robust even when journal fixed effects are included.

\begin{table}[H]
    \centering
    \caption{Prediction of Citations: Health versus Non-Health}\label{tab:predcitH}
\scalebox{0.91}{ \begin{tabular}{lcccccccc}
    \toprule
  &\multicolumn{4}{c}{\textsc{Unconditional}}&\multicolumn{4}{c}{\textsc{Conditional on Journal}}\\
\cmidrule(lr){2-5}\cmidrule(lr){6-9}
& (1) & (2) & (3) & (4) & (5) & (6) & (7) & (8) \\
\midrule
Health                   &     -0.4393***&     -0.4368***&     -0.4762***&     -0.4769***&     -0.0455   &     -0.0422   &     -0.0662   &     -0.0671   \\
                         &     (0.028)   &     (0.028)   &     (0.030)   &     (0.030)   &     (0.047)   &     (0.047)   &     (0.048)   &     (0.048)   \\
Novelty                  &      0.0585***&               &      0.1428***&               &      0.0742***&               &      0.1559***&               \\
                         &     (0.013)   &               &     (0.015)   &               &     (0.013)   &               &     (0.015)   &               \\
Health $\times$ Novelty &      0.0028   &               &     -0.0596*  &               &     -0.0134   &               &     -0.0718** &               \\
                         &     (0.029)   &               &     (0.032)   &               &     (0.028)   &               &     (0.031)   &               \\
Impact                   &               &      0.0770***&      0.1542***&               &               &      0.0608***&      0.1464***&               \\
                         &               &     (0.013)   &     (0.015)   &               &               &     (0.013)   &     (0.015)   &               \\
Health $\times$ Impact  &               &     -0.0452   &     -0.0914***&               &               &     -0.0276   &     -0.0813** &               \\
                         &               &     (0.030)   &     (0.033)   &               &               &     (0.029)   &     (0.032)   &               \\
Quality                  &               &               &               &      0.1505***&               &               &               &      0.1534***\\
                         &               &               &               &     (0.014)   &               &               &               &     (0.013)   \\
Health $\times$ Quality &               &               &               &     -0.0757***&               &               &               &     -0.0771***\\
                         &               &               &               &     (0.027)   &               &               &               &     (0.027)   \\
\midrule Journal F.E.    &               &               &               &               &\(\checkmark\)   &\(\checkmark\)   &\(\checkmark\)   &\(\checkmark\)   \\
\midrule
N                        &      35,208   &      35,208   &      35,208   &      35,208   &      35,208   &      35,208   &      35,208   &      35,208   \\

\bottomrule
    \end{tabular}}
\begin{minipage}{\textwidth}  
\scriptsize{}\textsc{Note.---}OLS regressions of year-normalised citation counts on the main analysis sample. Independent variables are measured in standard deviations. Standard errors in parentheses. * $p < 0.05$, ** $p < 0.01$, *** $p < 0.001$.
\end{minipage}
\end{table}

\section{Conclusion}\label{sec:Conclusion}

The evolving role of health economics within general economic research and publishing over the past 30 years marks a significant transformation from a niche field to a mainstream area of interest. Historical contributions from seminal scholars like \citet{arrow1963uncertainty}, \citet{newhouse1970toward}, and \citet{grossman1972concept} laid foundational stones, yet health economics rarely featured in prestigious general-interest economics journals. However, recent decades have witnessed a dramatic shift, with health economists increasingly contributing to and gaining recognition in broader economic discourse. This trend is vividly illustrated by the rising proportion of health economics papers in top-tier and general-interest journals, and the substantial growth in health economics research in related fields such as labor, development, and public economics.

Our analysis identifies several key factors behind the rising prominence of health economics in academic publishing. We utilize advanced bibliometric methods, leveraging natural language processing (NLP) techniques with RoBERTa to accurately classify health economics papers. Our findings suggest that the integration of health economics into mainstream economics is driven primarily by innovative, high-quality research rather than mere conformity to existing norms. By adapting \cite{kelly2021measuring}'s methods to evaluate the novelty, impact, and quality of academic papers, we demonstrate that health economics papers consistently exhibit higher quality, indicating a substantial contribution to the field's evolution. This shift not only underscores the field's growing importance but also highlights its potential to influence and enrich general economic research. As we expand on these findings in the full paper, we aim to provide a comprehensive understanding of the dynamic interplay between health economics and broader economic thought, charting the course for future research and policy implications.

\clearpage 
\doublespacing
\bibliographystyle{aer}
\bibliography{bibliometrics}

\clearpage

\clearpage 

\setcounter{section}{0}
\setcounter{page}{1}

\clearpage

\doublespacing

\appendix
\renewcommand \thepart{}  
\renewcommand \partname{} 

\addcontentsline{toc}{section}{Appendix} 
\part{Appendix}  
\parttoc 

\renewcommand*{\theHsection}{chY.\the\value{section}}

\renewcommand{\thefigure}{\Alph{section}.\arabic{figure}}
\renewcommand{\thetable}{\Alph{section}.\arabic{table}}

\section{Additional tables and figures}\label{sec:app}
\setcounter{figure}{0} 
\setcounter{table}{0}

\begin{figure}[H]
    \caption{Binned scatter plots of text-based measures on citations}\label{app:binscatters}
    \centering
    \begin{subfigure}{.32\textwidth}
		\centering
        \includegraphics[width=\textwidth]{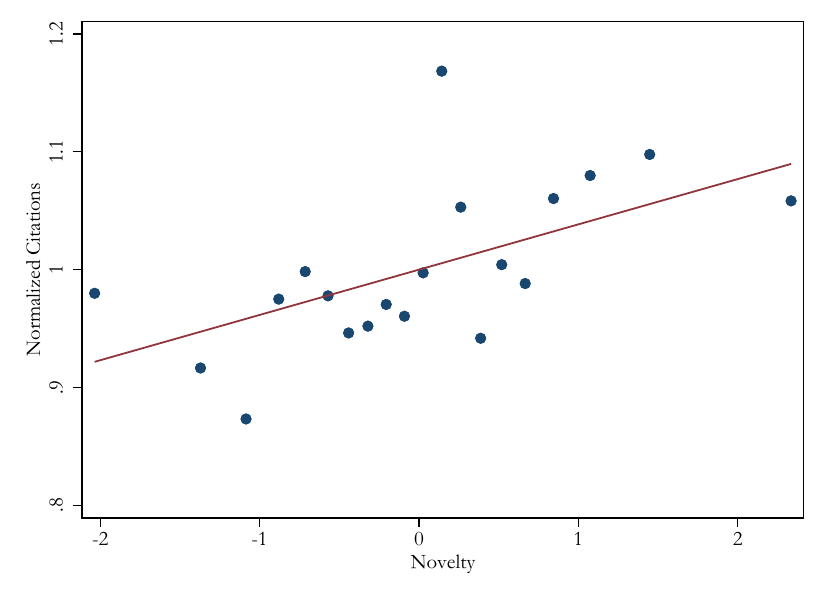}
		\label{}
	\end{subfigure}
    \begin{subfigure}{.32\textwidth}
		\centering
        \includegraphics[width=\textwidth]{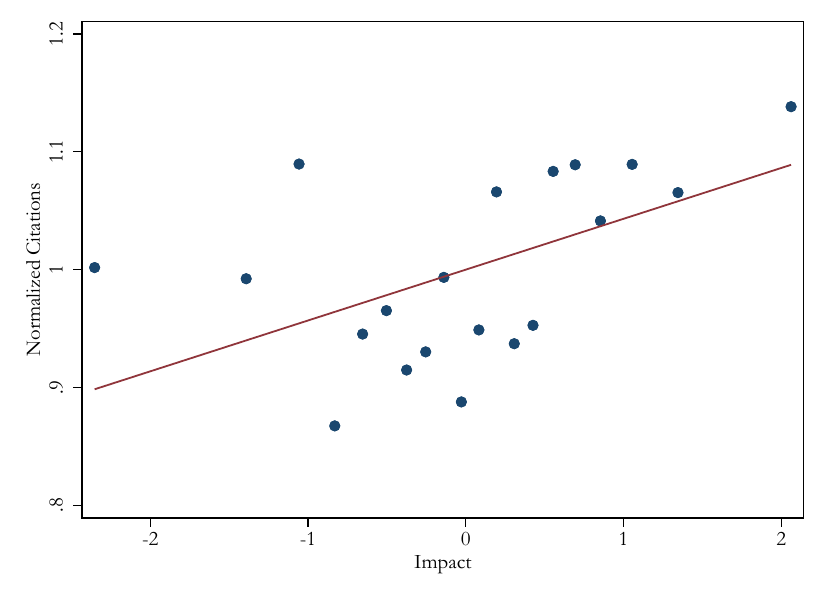}
		\label{}
	\end{subfigure}
     \begin{subfigure}{.32\textwidth}
		\centering
        \includegraphics[width=\textwidth]{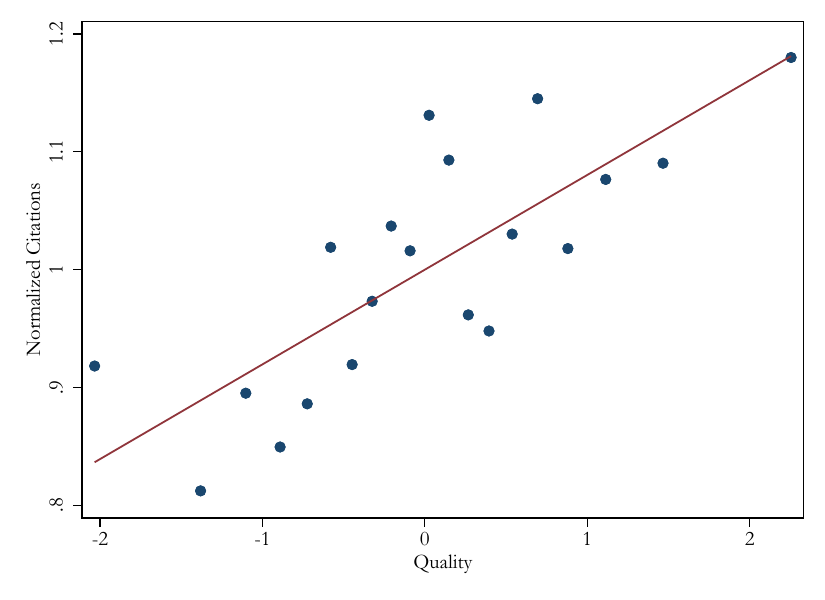}
		\label{}
	\end{subfigure}
\begin{minipage}{\textwidth}  
\scriptsize{}\textsc{Note.--- This Figure shows binned scatter plots with 20 bins of our text-based measures and normalized citations for each health economics paper in our sample. The regression fit in the plots represents the regressions reported in \ref{tab:predcit} Columns (1), (2), and (4), without journal fixed effects.}
\end{minipage}
\end{figure}

\begin{figure}[H]
    \caption{Binned scatter plots of text-based measures on citations with journal fixed effects}\label{app:binscatterswfe}
    \centering
    \begin{subfigure}{.32\textwidth}
		\centering
        \includegraphics[width=\textwidth]{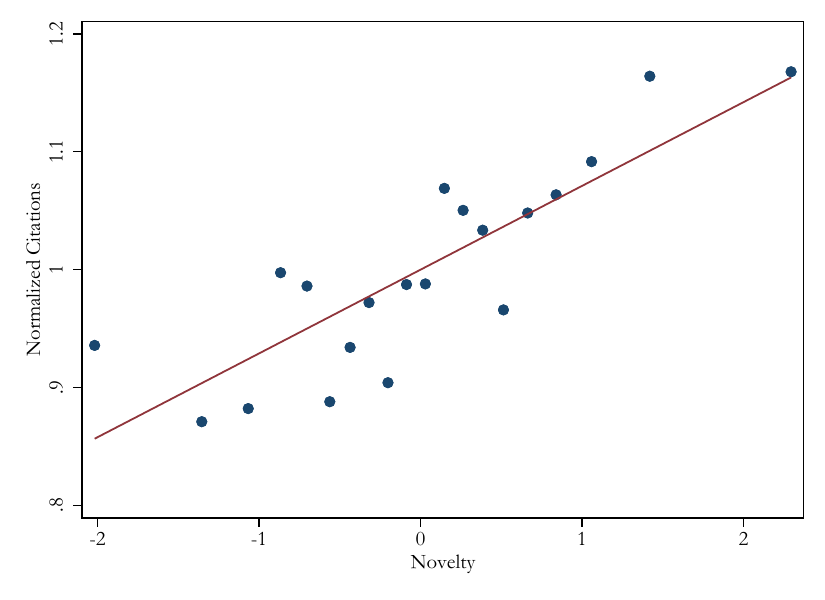}
		\label{}
	\end{subfigure}
    \begin{subfigure}{.32\textwidth}
		\centering
        \includegraphics[width=\textwidth]{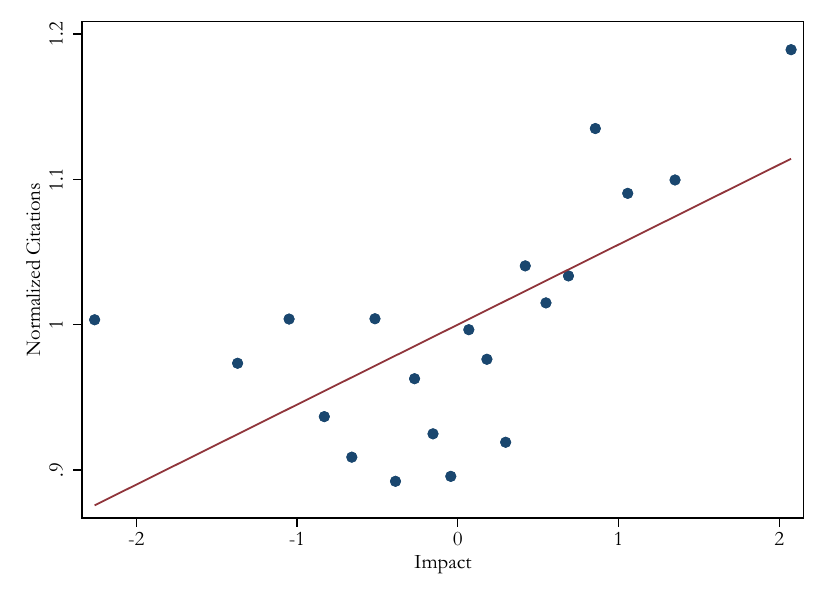}
		\label{}
	\end{subfigure}
     \begin{subfigure}{.32\textwidth}
		\centering
        \includegraphics[width=\textwidth]{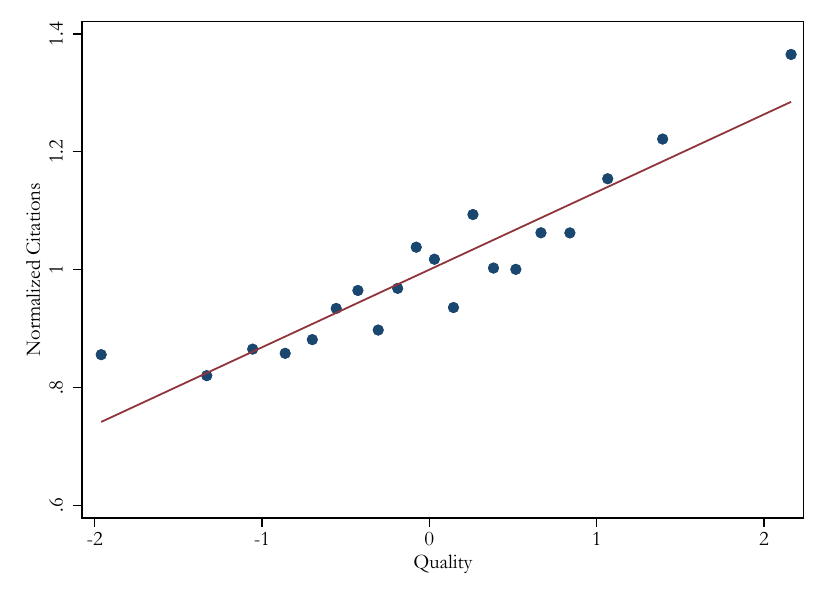}
		\label{}
	\end{subfigure}
\begin{minipage}{\textwidth}  
\scriptsize{}\textsc{Note.--- This Figure shows binned scatter plots with 20 bins of our text-based measures and normalized citations for each health economics paper in our sample. The regression fit in the plots represents the regressions reported in \ref{tab:predcit} Columns (5), (6), and (8), including journal fixed effects.}
\end{minipage}
\end{figure}

In Figures \ref{app:binscatters} and \ref{app:binscatterswfe}, we provide additional visualizations of the regressions shown in Table \ref{tab:predcit}.

\end{document}